\documentclass[12pt]{article}

\usepackage{amssymb,amsmath,epsfig,cite}

\normalbaselines

\begin{document}

\title{Trends in supersymmetric quantum mechanics\footnote{Dedicated to my dear friend and colleague V\'eronique Hussin}}

\author{{\Large David J. Fern\'andez C.} \\[10pt] \it Departamento de F\'{\i}sica, Cinvestav \\ \it  A.P. 
14-740, 07000 Ciudad de M\'exico \\ \it M\'exico}

\date{}

\maketitle

\begin{abstract}
Along the years, supersymmetric quantum mechanics (SUSY QM) has been used for studying solvable quantum 
potentials. It is the simplest method to build Hamiltonians with prescribed spectra in the spectral 
design. The key is to pair two Hamiltonians through a finite order differential operator. Some related 
subjects can be simply analyzed, as the algebras ruling both Hamiltonians and the associated coherent 
states. The technique has been applied also to periodic potentials, where the spectra consist of allowed 
and forbidden energy bands. In addition, a link with non-linear second order differential equations, and 
the possibility of generating some solutions, can be explored. Recent applications concern the study of 
Dirac electrons in graphene placed either in electric or magnetic fields, and the analysis of optical 
systems whose relevant equations are the same as those of SUSY QM. These issues will be reviewed briefly 
in this paper, trying to identify the most important subjects explored currently in the literature.
\end{abstract}

\maketitle


\section{Introduction}

The birth of supersymmetric quantum mechanics (SUSY QM) in 1981, as a toy model to illustrate the 
properties that systems involving both bosons and fermions have, was a breakthrough in the study of 
solvable quantum mechanical models \cite{wi81}. One of the reasons is that SUSY QM is tightly related to 
other approaches used in the past to address this kind of systems, {\it e.g.}, the factorization method, 
Darboux transformation and intertwining technique \cite{ih51,mi68,lrb90,rrr91,lr91,cks95,ju96,zc97,rc99,ro99,ba01,kks01,ro02,afhnns04,mr04,ks04,bbrr04,bs04,ac04,kh05,su05,ff05,do07,fe10,ai12,bsps14}. 

On the other hand, it is well known that the factorization method was introduced by Dirac in 1935, to 
derive algebraically the spectrum of the harmonic oscillator \cite{di35}. The next important advance was 
done by Schr\"odinger in 1940, who realized that the procedure can be also applied to the Coulomb 
potential \cite{sch40,sch41}. Later on, Infeld and his collaborators push forward the technique 
\cite{in41,hi48}, supplying a general classification scheme including most of the exactly solvable 
Schr\"odinger Hamiltonians known up to that time \cite{ih51}. As a consequence, the idea that the 
factorization methods was essentially exhausted started to spread among the scientific community.

However, in 1984 Mielnik proved that this belief was wrong, by generalizing simply the Infeld-Hull 
factorization method when he was seeking the most general first-order differential operators which 
factorize the harmonic oscillator Hamiltonian in a certain given order \cite{mi84}. The key point of 
his approach was that if the ordering of the generalized factorization operators is interchanged, then a 
new Hamiltonian is obtained which is intertwined with the oscillator one. 

It is worth to stress that Mielnik's work represented the next breakthrough in the development of the 
factorization method, since it opened the way to look for new solvable quantum potentials. In 
particular, this generalization was immediately applied to the Coulomb problem \cite{fe84}. Meanwhile, 
Andrianov's group \cite{abi84a,abi84b} and Nieto \cite{ni84} identified the links of the factorization 
method with Darboux transformation and supersymmetric quantum mechanics, respectively. In addition, 
Sukumar indicated the way to apply Mielnik's approach to arbitrary potentials and factorization 
energies \cite{su85a,su85b}, setting up the general framework where the factorization method would 
develop for the next decade \cite{mi84,fe84,su85a,lp86,bh86,bdh87,ad88,ar90,adv91,bdn92,ca95,fs96,jr97,as97,dr97,brsv98,mnn98,cmpr98,qv98,qv99,ek99,cr00,bmq01,qu03,rmc03,iknn06,kn08,qu08}. 

Let us mention that up to the year 1993 the factorization operators, which at the same time are 
intertwining operators in this case, were first order differential ones. A natural generalization, 
pursued by Andrianov and collaborators \cite{ais93,aicd95}, consists in taking the intertwining 
operators of order greater than one. This proposal was important, since it helped to circumvent the 
restriction of the first-order method, that only the energy of the initial ground state can be modified.
Moreover, it made clear that the key of the generalization is the analysis of the intertwining relation
rather than the factorized expressions. Let us note also that in 1995 Bagrov and Samsonov explored the 
same technique in a different but complementary way \cite{bs95}. 

Our group got back to the subject in 1997 \cite{fe97,fgn98,fhm98,fr01,crf01}, although some works related with the 
method had been done previously \cite{fno96}. In particular, several physically interesting 
potentials were addressed through this technique, as the standard harmonic oscillator 
\cite{mi84,fgn98,fhm98}, the radial oscillator and Coulomb potentials \cite{fe84,fno96,fs05}, among 
others \cite{cf07,cf08,fm18}. In addition, the coherent 
states associated with the SUSY partners of the harmonic oscillator were explored 
\cite{fhn94,fnr95,fh99,bcf14}, and similar works dealing with more general one-dimensional Hamiltonians  
were done \cite{fhr07,fhm18}. Another important contribution has to do with the determination of the 
general systems ruled by polynomial Heisenberg algebras and the study of particular realizations based 
on the SUSY partners of the oscillator \cite{fh99,fnn04,cfnn04,mn08,cdf18}. The complex SUSY transformations 
involving either real or complex factorization `energies' were as well implemented 
\cite{fmr03,fr08,bf11b,be12,fg15}. In addition, the analysis of the confluent algorithm, the degenerate case in 
which all the factorization energies tend to a single one, was also elaborated \cite{fs05,mnr00,fs03,fs11,bff12,cs14,cs15a,cs15b,be16,cs17}. 
The SUSY techniques for exactly solvable periodic potentials, as the Lam\'e and associated Lam\'e 
potentials, have been as well explored \cite{fnn00,nnf00,fmrs02a,fmrs02b,fg05,gin06,fg07,fg10}.

Some other groups have addressed the same subjects through different viewpoints, e.g., the $N$-fold 
supersymmetry by Tanaka and collaborators \cite{ast01,st02,ta03,gt04,bt09}, the hidden nonlinear 
supersymmetry by Plyushchay et al \cite{lp03,pl04,cnp07,cjnp08,cjp08}, among others.

Specially important is the connection of SUSY QM with non-linear second-order ordinary differential 
equations, as KdV and Painlev\'e IV and V equations, as well as the possibility of designing algorithms 
to generate some of their solutions 
\cite{dek92,vs93,ad94,srk97,acin00,fnn04,cfnn04,mn08,ma09a,ma09b,bf11a,bf11b,bf13,bf14,fm14,bcf14,fg15,fm16,bfn16,mq16,no18,cggm18}.

Another relevant subject related to SUSY QM is the so-called exceptional orthogonal polynomials (EOP) \cite{cprs08,os09a,os09b,gkm10,os11,qu11a,qu11b,gkm12,gkm13,mq13,mq14,ggm14a,ggm14b,sr14,ggm18}. 
In fact, it seems that most of these new polynomials appear quite naturally when the seed solutions 
which are employed reduce to polynomial solutions of the initial stationary Schr\"odinger equation 
\cite{gkm13}.  

Recently, the SUSY methods started to be used also in the study of Dirac electrons in graphene and some 
of its allotropes, when external electric or magnetic fields are applied 
\cite{knn09,mtm11,jp12,pr12,jknt13,qs13,mf14,knt15,sr17,chrv18,jss18}. 
It is worth to mention as well some systems in optics, since there is a well known correspondence between 
Schr\"odinger equation and Maxwell equations in the paraxial approximation, which makes that the SUSY 
methods can be applied directly in some areas of optics \cite{cw94,mhc13,zrfm14,mi14,dllrc15,cg17,mwlf18,cj18}.

As we can see, the number of physical systems which are related with supersymmetric quantum mechanics is 
large enough to justify the writing of a new review paper, in which we will present the recent advances 
in the subject. If the reader is looking for books and previous review papers addressing SUSY QM from an
inductive viewpoint, we recommend references \cite{rrr91,lr91,cks95,ju96,zc97,rc99,ro99,ba01,kks01,ro02,afhnns04,mr04,ks04,bbrr04,bs04,ac04,kh05,su05,ff05,do07,fe10,ai12,bsps14}.

\section{Supersymmetric quantum mechanics}

In this section we shall present axiomatically the supersymetric quantum mechanics, as a tool for 
generating solvable potentials $\widetilde V(x)$ departing from a given initial one $V(x)$.

The supersymmetry algebra with two generators introduced by Witten in 1981 \cite{wi81}
\begin{eqnarray}
&& [Q_i, H_{\rm ss}]=0, \quad \{ Q_i,Q_j \} = \delta_{ij} H_{\rm ss}, \quad i,j=1,2 , \label{susyalg}
\end{eqnarray}
when realized in the following way
\begin{eqnarray}
&& Q_1 =  \frac{Q^+ + Q}{\sqrt{2}}, \qquad Q_2 = \frac{Q^+ - Q}{i\sqrt{2}}, \\
&& Q = \left(\begin{matrix} 0 & 0 \\ B & 0 \end{matrix} \right), \quad Q^+ =
\left(\begin{matrix} 0 & B^+ \\ 0 & 0 \end{matrix}
\right), \\
&& H_{\rm ss} = \{ Q, Q^+ \} = \left(\begin{matrix} B^+ B & 0 \\ 0 & B B^+ \end{matrix} 
\right) ,
\end{eqnarray}
is called supersymmetric quantum mechanics, where $H_{\rm ss}$ is the supersymmetric Hamiltonian while 
$Q_1, \ Q_2$ are the supercharges. The $k$th order differential operators $B, \ B^+$ intertwine two 
Schr\"odinger Hamiltonians 
\begin{eqnarray}\label{HandHt}
&&
\widetilde H = - \frac12 \frac{d^2}{dx^2} + \widetilde V(x),  \qquad
H = - \frac12 \frac{d^2}{dx^2} + V(x), 
\end{eqnarray}
in the way
\begin{eqnarray}\label{intertwining}
&& \widetilde H B^+ = B^+ H, \qquad H B = B \widetilde H.
\end{eqnarray}
There is a natural link with the {\it factorization method}, since the following relations are 
fulfilled:
\begin{eqnarray}
&& B^+ B = \prod\limits_{j=1}^{k} (\widetilde H - \epsilon_j), \qquad
B B^+ = \prod\limits_{j=1}^{k} (H - \epsilon_j),
\end{eqnarray}
where $\epsilon_j, j=1,\dots,k$ are $k$ {\it factorization energies} associated to $k$ {\it seed 
solutions} required to implement the intertwining (see equations~(\ref{HandHt}-\ref{intertwining}) and
Sections~\ref{sst} and \ref{cst} below). Taking into account these expressions, it turns out that the 
supersymmetric Hamiltonian $H_{\rm ss}$ is a polynomial of degree $k$th in the diagonal matrix operator 
$H_{\rm p}$ which involves the two Schr\"odinger Hamiltonians $H$ and $\widetilde H$ as follows
\begin{eqnarray}
&& H_{\rm ss} =  \prod\limits_{j=1}^{k} (H_{\rm p} - \epsilon_j), \qquad
H_{\rm p}  =  \left(
\begin{matrix}
\widetilde H & 0 \\
0 & H
\end{matrix}
\right).
\end{eqnarray}
In particular, if $k=1$ the standard (first-order) supersymmetric quantum mechanics is recovered, for 
which $H_{\rm ss}$ is a first degree polynomial in $H_{\rm p}$, $H_{\rm ss} = H_{\rm p} - \epsilon_1$. 
For $k>1$, however, we will arrive to the so-called higher-order supersymmetric quantum mechanics, in 
which $H_{\rm ss}$ is a polynomial of degree greater than one in $H_{\rm p}$ (see for example 
\cite{ff05}).

\subsection{Standard SUSY transformations}\label{sst}

Let us suppose now that we select $k$ solutions $u_j$ of the initial stationary Schr\"odinger equation 
for $k$ {\it different} factorization energies $\epsilon_j, j=1,\dots,k$,
\begin{eqnarray}
& H u_j = \epsilon_j u_j,
\end{eqnarray}
which are called {\it seed solutions}. From them we implement the intertwining transformation of equation 
(\ref{intertwining}), leading to a new potential $\widetilde V(x)$ which is expressed in terms of the 
initial potential and the seed solutions as follows:
\begin{eqnarray}
& \widetilde V(x) = V(x) - [\log W(u_1, \dots,u_k)]'' , \label{nVk}
\end{eqnarray}
where $W(u_1, \dots,u_k)$ denotes the Wronskian of $u_j, j=1,\dots,k$. The eigenfunctions $\widetilde\psi_n$ 
and eigenvalues $E_n$ of $\widetilde H$ are obtained from the corresponding ones of $H$, $\psi_n$ and $E_n$, 
as follows:
\begin{eqnarray}
& \widetilde\psi_n = & \frac{B^+ \psi_n}{\sqrt{(E_n - \epsilon_1) \cdots (E_n - \epsilon_k)}} \propto
\frac{W(u_1,\dots,u_k,\psi_n)}{W(u_1,\dots,u_k)}. \label{neign}
\end{eqnarray}
Moreover, $\widetilde H$ could have additional eigenfunctions $\widetilde\psi_{\epsilon_j}$ for some of 
the factorization energies $\epsilon_j$ (at most $k$, depending of either they fulfill or not the 
required boundary conditions) which are given by:
\begin{eqnarray}
& \widetilde\psi_{\epsilon_j} \propto & \frac{W(u_1,\dots,u_{j-1},u_{j+1},\dots, u_k)}{W(u_1,\dots,u_k)}.
\label{neigep}
\end{eqnarray}

We can conclude that, given the initial potential $V(x)$, its eigenfunctions $\psi_n$, eigenvalues $E_n$ and 
the $k$ chosen seed solutions  $u_j, j=1,\dots,k$, it is possible to generate algorithmically its $k$th 
order SUSY partner potential $\widetilde V(x)$ as well as the associated eigenfunctions and eigenvalues 
through expressions (\ref{nVk}-\ref{neigep}).

It is important to stress that the seed solutions must be carefully chosen in order that the new 
potential will not have singularities additional to those of the initial potential $V(x)$. When this 
happens, we say that the transformation is {\it non-singular}. If the initial potential is real, and we 
require the same for the final potential, then there are some criteria for choosing the real seed 
solutions $u_j$ according to their number of nodes, which also depend on the values taken by the 
associated factorization energies $\epsilon_j$ (see for example \cite{ff05}). Although non-exhaustive, let 
us report next a list of some important criteria, which will make the final potential $\widetilde V(x)$ 
to be real and without any extra singularity with respect to $V(x)$.

\begin{itemize}

\item[-] If $k=1$ (first-order SUSY QM), the factorization energy $\epsilon_1$ must belong to the 
infinite energy gap $\epsilon_1 < E_0$ in order that $u_1$ could be nodeless inside the $x$-domain of 
the problem, where $E_0$ is the ground state energy of $H$. Moreover, since in this $\epsilon_1$-domain 
the seed solution $u_1$ could have either one node or none, then we additionally require to identify the
right nodeless solution. With these conditions, the transformation will be non-singular and the spectrum 
of the new Hamiltonian $\widetilde H$ will have an extra level $\epsilon_1$ with respect to $H$ 
(creation of a new level). Note that also it is possible to select the seed solution with a node at one 
of the edges of the $x$-domain, thus the SUSY	transformation will be still non-singular but the 
factorization energy $\epsilon_1$ will not belong to the spectrum of $\widetilde H$ (isospectral 
transformation).

\item[-] If $k=1$, $\epsilon_1 = E_0$ and $u_1 = \psi_0$ (the seed solution is the ground state, which 
has one node at each edge of the $x$-domain), then the SUSY transformation will be non-singular and the 
spectrum of the new Hamiltonian will not have the level $E_0$ (deletion of one level).

\item[-] If $k=2$ (standard second-order SUSY QM), first of all both $\epsilon_1$ and $\epsilon_2$ must 
belong to the same energy gap, either to the infinite one below $E_0$ or to a finite gap defined by two 
neighbor energy levels $(E_m, E_{m+1})$. Let us order the two factorization energies in the way 
$\epsilon_2<\epsilon_1$. In order that the wronskian of $u_1$ and $u_2$ would be nodeless, the seed 
solution $u_2$ associated to the lower factorization energy $\epsilon_2$ should have one extra node with 
respect to the solution $u_1$ associated to the higher factorization energy $\epsilon_1$ \cite{ff05}. In 
particular, in the infinite gap $u_2$ should have one node and $u_1$ should be nodeless. On the other 
hand, when both factorization energies are in the finite gap $(E_m, E_{m+1})$ the seed solutions $u_2$ 
and $u_1$ should have $m+2$ and $m+1$ nodes respectively. In both cases the spectrum of the new 
Hamiltonian will contain two extra eigenvalues $\epsilon_1, \epsilon_2$ (creation of two levels). 
Moreover, the seed solutions can be chosen such that the transformation is still non-singular but either 
$\epsilon_1$, $\epsilon_2$ or both will not belong to the spectrum of $\widetilde H$ (either creation of 
one new level or isospectral transformation).

\item[-] If $k=2$, $\epsilon_2=E_m$, $u_2 = \psi_m$, $\epsilon_1=E_{m+1}$, $u_1 = \psi_{m+1}$, then the
SUSY transformation will be non-singular and the spectrum of the new Hamiltonian will not have the two 
levels $E_m, E_{m+1}$ (deletion of two levels).

\item[-] If $k>2$, the corresponding non-singular SUSY transformation can be expressed as the product of 
a certain number of first and second-order SUSY transformations, each one having to be consistent with 
any of the previous criteria to be non-singular. 

\end{itemize}

\subsection{Confluent SUSY transformations}\label{cst}

An important degenerate case of the SUSY transformation for $k\geq 2$ appears when all the factorization 
energies $\epsilon_j, j=1,\dots,k$ tend to a fixed single value $\epsilon_1$ 
\cite{mnr00,fs03,fs05,fs11,bff12,sc13,cs14,cs15a,cs15b,gq15,be16,kr16,cs17}. Let us note that the 
expression for the new potential of equation~(\ref{nVk}) is still valid, but the seed solutions have to 
be changed if non-trivial modifications in the new potential are going to appear. In fact, the seed 
solutions $u_j, j=1,\dots,k$ instead of being just normal eigenfunctions of 
$H$ should generate a Jordan chain of generalized eigenfunctions for $H$ and $\epsilon_1$ as follows:
\begin{eqnarray}
& (H-\epsilon_1)u_1 = & \hskip-0.2cm 0 , \label{chain-homo} \\ 
& (H-\epsilon_1)u_2 = & \hskip-0.2cm u_1,\\
& \vdots & \nonumber \\
& (H-\epsilon_1)u_{k} = & \hskip-0.2cm u_{k-1}.
\end{eqnarray}

First let us assume that the seed solution $u_1$ satisfying equation (\ref{chain-homo}) is given, then 
we need to find the general solution for $u_j, j=2,\dots,k$ (precisely in that order!) in terms of 
$u_1$. There are two methods essentially different to determine such a general solution: the first one 
is known as integral method, in which through the technique of variation of parameters one simplifies 
each inhomogeneous equation in the chain and when integrating the resulting equation every solution 
$u_j$ is found. In fact, by applying this procedure the solution to the inhomogeneous equations
\begin{eqnarray}
(H-\epsilon_1)u_j = u_{j-1}, \quad j=2,\dots,k,
\end{eqnarray}
is given by
\begin{eqnarray}
u_j(x) & = & -2 \, u_1(x) \, v_j(x), \label{chainj1} \\
v_j(x) & = & v_j(x_0) + \int_{x_0}^{x} \frac{w_j(y)}{u_1^2(y)} dy, \label{chainj2} \\
w_j(x) & = & w_j(x_0) + \int_{x_0}^{x} u_1(z) \, u_{j-1}(z) dz, \label{chainj3}
\end{eqnarray}
where $x_0$ is a point in the initial domain of the problem. Thus, equation~(\ref{chainj3}) with $j=2$ 
determines $w_2$, by inserting then this result in equation~(\ref{chainj2}) with $j=2$ we find $v_2$ 
which in turn fixes $u_2$ through equation~(\ref{chainj1}) \cite{fs05}. By using then this expression 
for $u_2$ it is found $w_3$ through equation~(\ref{chainj3}) and then $v_3$ and $u_3$ by means of 
equations~(\ref{chainj2}) and (\ref{chainj1}) respectively \cite{fs11}. We continue this process to find 
at the end the expression for $u_k$, and then we insert all the $u_j, j=1,\dots,k$ in 
equation~(\ref{nVk}) in order to obtain the new potential \cite{sc13}.

An alternative is the so-called differential method, in which one identifies in a clever way (through 
parametric differentiation with respect to the factorization energy $\epsilon_1$) one particular 
solution for each inhomogeneous equation of the chain \cite{bff12,be16}. Is is straightforward then to 
find the general solution for each $u_j, j=2,\dots,k$. Instead of supplying the resulting formulas for 
arbitrary $k>1$, let us derive the results just for the simplest case with $k=2$.

\subsubsection{Confluent second-order SUSY QM}

For $k=2$ we just need to solve the following system of equations:
\begin{eqnarray}
& (H-\epsilon_1)u_1 = & \hskip-0.2cm 0 , \label{chain-homok21}\\ 
& (H-\epsilon_1)u_2 = & \hskip-0.2cm u_1. \label{chain-homok22}
\end{eqnarray}
The result for the integral method in this case is achieved by making $k=2$ in 
equations~(\ref{chainj1}-\ref{chainj3}), which leads to \cite{fs05}:
\begin{eqnarray}
u_2(x) & = & -2 \, u_1(x) \, v_2(x), \label{2confla} \\
v_2(x) & = & v_2(x_0) + \int_{x_0}^{x} \frac{w_2(y)}{u_1^2(y)} dy, \label{2conflb} \\
w_2(x) & = & w_2(x_0) + \int_{x_0}^x u_1^2(y) dy. \label{2conflc}
\end{eqnarray}
Thus we obtain:
\begin{eqnarray}
W(u_1,u_2)= -2 \, w_2(x). \label{confwrons}
\end{eqnarray}
Up to a constant factor, this is the well known formula generated for the first time in \cite{fs03}, 
which will induce non-trivial modifications in the new potential $\widetilde V(x)$ (see 
equation~(\ref{nVk})).

Let us solve now the system of equations~(\ref{chain-homok21}-\ref{chain-homok22}) through the 
differential method \cite{bff12}. If we derive equation~(\ref{chain-homok21}) with respect to 
$\epsilon_1$, assuming that the Hamiltonian $H$ does not depend explicitly on $\epsilon_1$, we obtain a 
particular solution of the inhomogeneous equation~(\ref{chain-homok22}), namely,
\begin{eqnarray}
(H-\epsilon_1) \frac{\partial u_1}{\partial\epsilon_1} = u_1. \label{diff2cp}
\end{eqnarray}
Thus, the general solution for $u_2$ we were looking for becomes:
\begin{eqnarray}
&& u_2(x) = c_2 \, u_1 + d_2 \, u_1\int_{x_0}^x \frac{dy}{u_1^2(y)} + \frac{\partial u_1}{\partial\epsilon_1}.
\label{gsdiff2}
\end{eqnarray}
Hence:
\begin{eqnarray}
W(u_1,u_2) = d_2 + W\left(u_1,\frac{\partial u_1}{\partial\epsilon_1}\right).
\end{eqnarray}

Let us note that both methods have advantages and disadvantages, as compared with each other. For 
instance, in the integral method often it is hard to find explicit analytic solutions for the involved 
integrals, then in such cases we can try to use the differential method. However, for numerical 
calculation of the new potential it is simple and straightforward to use the integral formulas. On the 
other hand, there are not many potentials for which we can calculate in a simple way the corresponding 
derivative with respect to the factorization energy. At the end both methods turn out to be 
complementary to each other. A final remark has to be done: the family of new potentials generated 
through both algorithms (the integral and differential one) is the same, but if we want to generate a 
specific member of the family through both methods we need to be sure that we are using the same pair of 
seed solutions $u_1, u_2$. In practice, given $u_1, u_2$, with $u_2$ generated for example through the 
integral method (which means that we have fixed the constants $v_2(x_0)$ and $w_2(x_0)$ of 
equations~(\ref{2conflb}, \ref{2conflc})) we have to look for the appropriate coefficients $c_2$ and 
$d_2$ of equation~(\ref{gsdiff2}) in order to guarantee that the same seed solution $u_2$ is going to be 
used for the differential algorithm (see the discusion at \cite{cs15b}).

As in the non-confluent SUSY approach, once again we have to choose carefully the seed solution $u_1$ in 
order that the new potential will not have extra singularities with respect to $V(x)$. In the case of 
the second-order confluent algorithm, the way of selecting such a seed solution is the following 
\cite{fs03}:

\begin{itemize}

\item[-] In the first place $u_1$ must vanishes at one the two edges of the $x$-domain. If this happens, 
then there will be some domain of the parameter $w_2(x_0)$ for which the key function $w_2$ of 
equation~(\ref{2conflc}) will not have any node.
 
\item[-] The above requirement can be satisfied, in principle, by seed solutions $u_1$ associated to any 
real factorization energy, thus we can create an energy level at any place on the energy axis.

\item[-] In particular, any eigenfunction of $H$ satisfies the conditions to produce non-singular 
confluent second-order SUSY transformations, and the corresponding energy eigenvalue can be also kept in 
the spectrum of the new Hamiltonian (isospectral transformations).

\item[-] When an eigenfunction of $H$ is used, a zero for $w_2$ could appear at one of the edges of the 
$x$ domain. In such a case, the SUSY transformation stays non-singular,  but the corresponding 
eigenvalue will disappear from the spectrum of $\widetilde H$ (deletion of one level).

\end{itemize}

\section{SUSY QM and exactly solvable potentials}

The methods discussed previously can be used to generate, from an exactly solvable potential, plenty of 
new exactly solvable Hamiltonians with spectra quite similar to the initial one. In this section we will 
employ the harmonic oscillator to illustrate the technique. Although in this case the spectrum consists 
of an infinite number of non-degenerate discrete energy levels, the method works as well for 
Hamiltonians with mixed spectrum (discrete and continuous) or even when there is just a continuous 
one (see e.g. \cite{ra04}). This is what happens for periodic potentials 
\cite{fnn00,nnf00,fmrs02a,fmrs02b,fg05,gin06,fg07,fg10}, where the spectrum consists of allowed energy 
bands separated by forbidden gaps. Moreover, the technique has been applied also to a very special 
system whose spectrum is the full real line, with each level being doubly degenerate: the so-called 
repulsive oscillator \cite{bf13}.

\subsection{Harmonic oscillator}

The harmonic oscillator potential is given by:
\begin{eqnarray}
V(x) = \frac{x^2}2.
\end{eqnarray}
In order to apply the SUSY methods, it is required to find the general solution $u(x)$ of the 
stationary Schr\"odinger equation for an arbitrary factorization energy $\epsilon$:
\begin{eqnarray}
&& -\frac12 u''(x) + \frac{x^2}2 u(x) = \epsilon \, u(x).
\end{eqnarray}
Up to a constant factor, the general solution to this equation is a linear combination (characterized by
the parameter $\nu$) of an even an odd linearly independent solutions, given by \cite{fh99}:
\begin{eqnarray}
u(x) & = & e^{-\frac{x^2}2}\left[{}_1F_1\left(\frac{1-2\epsilon}4,\frac12;x^2\right) + 2 \nu \frac{\Gamma(\frac{3-2\epsilon}4)}{\Gamma(\frac{1-2\epsilon}4)} \, x \, {}_1F_1\left(\frac{3-2\epsilon}4,\frac32;x^2\right)\right] \nonumber \\
& = & e^{\frac{x^2}2}\left[{}_1F_1\left(\frac{1+2\epsilon}4,\frac12;-x^2\right) + 2 \nu \frac{\Gamma(\frac{3-2\epsilon}4)}{\Gamma(\frac{1-2\epsilon}4)} \, x \, {}_1F_1\left(\frac{3+2\epsilon}4,\frac32;-x^2\right)\right]. \label{gsseho}
\end{eqnarray}

In order to produce non-singular SUSY transformations we need to know the number of nodes that $u$ has, 
according to the position of the parameter $\epsilon$ on the energy axis. Let us note first of all that, 
if $\epsilon$ is any real number, $u$ will have an even number of nodes for $|\nu|<1$ while this number 
will be odd for $|\nu|>1$. This implies that, when $\epsilon$ is in the infinite energy gap 
$\epsilon < E_0$, this solution will have one node for $|\nu|>1$ and it will be nodeless for $|\nu|<1$. 
On the other hand, if $E_m<\epsilon<E_{m+1}$ with $m$ even, then $u$ will have $m+2$ nodes for 
$|\nu|<1$ and it will have $m+1$ nodes for $|\nu|>1$, while for odd $m$ it will have $m+2$ and $m+1$ 
nodes for $|\nu|>1$ and $|\nu|<1$, respectively. 

Now, although the SUSY methods can supply an infinity of new exactly solvable potentials, their 
expressions become in general to long to be explicitly reported. The simplest formulas appear when 
the factorization energies become either some of the eigenvalues $E_n = n+\frac12, n=0,1,\dots$ of $H$ 
or some other special values, defined by the sequence ${\cal E}_m = - (m+\frac12), m=0,1,\dots$ In both 
cases it is possible to reduce the Schr\"odinger solution $u$ to the product of one exponential factor 
$e^{\pm x^2/2}$ times a Hermite polynomial, either of a real variable when one of the $E_n$ is taken or 
of an imaginary one when any of the ${\cal E}_m$ is chosen \cite{fh99}. We supply next some explicit 
expressions for exactly solvable potentials, generated through the SUSY methods for such special values 
of the factorization energies. Let us note that we have sticked strictly to the criteria pointed out at 
section 2.1 for producing non-singular SUSY transformations on the full real line. It is just for the 
first-order transformation that we have employed one general solution to show explicitly the simplest 
family of exactly solvable potential generated through SUSY QM.

\subsubsection{First-order SUSY partners of the oscillator}

For $k=1$, $\epsilon_1=-\frac12$, $|\nu_1|<1$ it is obtained (see also \cite{mi84}): 
\begin{eqnarray}
\widetilde V(x) = \frac{x^2}2 - \left(\frac{2\,\nu_1 \, e^{-x^2}}{\sqrt{\pi}\,[1+\nu_1\,{\rm erf}(x)]} 
\right)' - 1, \label{1susy-m12}
\end{eqnarray}
where ${\rm erf}(x)$ is the error function.

For $k=1$, $\epsilon_1=-\frac52$, $\nu_1=0$ we get:
\begin{eqnarray}
\widetilde V(x) = \frac{x^2}2 - \left(\frac{4x}{2x^2+1} \right)' - 1. \label{1susy-m52}
\end{eqnarray}

For $k=1$, $\epsilon_1=-\frac92$, $\nu_1=0$ it is obtained:
\begin{eqnarray}
\widetilde V(x) = \frac{x^2}2 - \left[\frac{8x(2x^2+3)}{4x^4+12x^2+3} \right]' - 1. \label{1susy-m92}
\end{eqnarray}

Let us note that in all these three cases the spectrum of the new Hamiltonian $\widetilde H$, besides 
having the eigenvalues of $H$, will contain also a new energy level at $\epsilon_1$.

\subsubsection{Second-order SUSY partners of the oscillator}

For $k=2$, $\epsilon_1=-\frac52$, $\nu_1=0$, $\epsilon_2=-\frac72$, $\nu_2\rightarrow \infty$ it is 
obtained:
\begin{eqnarray}
\widetilde V(x) = \frac{x^2}2 - \left(\frac{16x^3}{4x^4+3} \right)' - 2. \label{2susy-m5272}
\end{eqnarray}

For $k=2$, $\epsilon_1=-\frac92$, $\nu_1=0$, $\epsilon_2=-\frac{11}2$, $\nu_2\rightarrow \infty$ we get:
\begin{eqnarray}
\widetilde V(x) = \frac{x^2}2 - \left[\frac{32x^3(4x^4+12x^2+15)}{16x^8+64x^6+120x^4+45} \right]' - 2.
\label{2susy-m92112}
\end{eqnarray}

For $k=2$, $\epsilon_1=-\frac52$, $\nu_1=0$, $\epsilon_2=-\frac{11}2$, $\nu_2\rightarrow \infty$ it is 
obtained:
\begin{eqnarray}
\widetilde V(x) = \frac{x^2}2 - \left[\frac{4x(12x^4+20x^2+5)}{8x^6+20x^4+10x^2+5} \right]' - 2.
\label{2susy-m5292}
\end{eqnarray}
Once again, in all these cases the spectrum of the new Hamiltonian $\widetilde H$ will have two new 
levels at $\epsilon_1, \ \epsilon_2$, besides the eigenvalues $E_n$ of $H$. 

On the other hand, when deleting two neighbor energy levels of $H$ in order to create $\widetilde H$ we 
could obtain again some of the potentials reported above, up to an energy shift to align the 
corresponding energy levels (see e.g. \cite{ef16}). For instance, if we delete the first and second 
excited states of $H$ we recover the potential given in equation~(\ref{1susy-m52}), if we delete the 
second and third excited states we get again the potential in equation~(\ref{2susy-m5272}). Let us 
generate now a new potential by deleting the third and fourth excited states, which leads to:
\begin{eqnarray}
\widetilde V(x) = \frac{x^2}2 - \left[\frac{12x(4x^4-4x^2+3)}{8x^6-12x^4+18x^2+9} \right]' + 2.
\label{2susy-p7292}
\end{eqnarray}
Note that the corresponding Hamiltonian $\widetilde H$ will not have the levels $E_3=7/2, \ E_4=9/2$.

In order to present some potentials obtained through the confluent second-order SUSY QM, let us use once 
again the eigenstates of $H$. If the ground state is taken to implement the transformation, it is 
generated the same family of potentials of equation~(\ref{1susy-m12}). However, if the first excited 
state is employed, the following one-parameter family of potentials isospectral to the oscillator is 
gotten (see equations~(\ref{nVk},\ref{2conflc},\ref{confwrons})):
\begin{eqnarray}
\widetilde V(x) = \frac{x^2}2 - \left[\frac{4x^2}{\sqrt{\pi} (2 b_2+1) e^{x^2}+\sqrt{\pi} 
e^{x^2} \text{erf}(x)-2 x}\right]',
\label{2susy-c32}
\end{eqnarray}
where $b_2\equiv w_2(-\infty)$. For $b_2>0$ the new Hamiltonian $\widetilde H$ is isospectral to $H$. 
However, if $b_2=0$ the level $E_1$ will disappear from the spectrum of $\widetilde H$. 

Let us note that if a general eigenfunction $\psi_n(x)$ of $H$ is used to perform the confluent 
second-order transformation, an explicit expression for the key function $w_2(x)$ has been obtained, 
which will induce non-trivial modifications in the new potential \cite{fs03}.

\section{Algebraic structures of $H$, $\widetilde H$ and coherent states}

In this section we are going to analyze the kind of algebra that the SUSY partner Hamiltonian 
$\widetilde H$ will inherit from the initial one $H$. We are going to suppose that $H$ has an algebraic 
structure general enough to include the most important one-dimensional Hamiltonians appearing currently 
in the literature, as the harmonic oscillator \cite{fhr07}.

\subsection{Algebraic structure of $H$}

Let us suppose that the initial Schr\"odinger Hamiltonian $H$ has an infinite discrete spectrum whose 
non-degenerate energy levels $E_n,\, n=0,1,\dots$ are ordered as usual, $E_n<E_{n+1}$. Moreover, there is 
an explicit funcional dependence between the eigenvalues $E_n$ and the index $n$, i.e., $E_n = E(n)$, 
where $E(n)$ is well defined on the non-negative integers. For example, for the harmonic oscillator it 
turns out that $E(n) = n + \frac12$. In this section we will use Dirac notation, so that the eigenstates 
and eigenvalues satisfy:
\begin{eqnarray}
&& H \vert\psi_n\rangle = E_n \vert\psi_n\rangle, \quad n=0,1,\dots
\end{eqnarray}
The number operator $N$ is now introduced as
\begin{eqnarray}
&& N\vert\psi_n\rangle = n \vert\psi_n\rangle .
\end{eqnarray}

It can be defined now a pair of ladder operators of the system through
\begin{eqnarray}
&& a^- \vert\psi_n\rangle = r(n) \vert\psi_{n-1}\rangle, \\ 
&& a^+ \vert\psi_n\rangle = r^*(n+1)\vert\psi_{n+1}\rangle, \\ 
&& r(n) = e^{i\tau(E_{n}-E_{n-1})} \ \sqrt{E_{n}-E_0}, \quad \tau \in {\mathbb R}, \label{phfacal}
\end{eqnarray}
where $r^*(n)$ denotes the complex conjugate of $r(n)$. Thus, the {\it intrinsic algebra} of the system 
is defined by:
\begin{eqnarray}
&& [N,a^\pm] = \pm  a^\pm, \\
&& a^+ a^- = E(N) - E_0, \\ 
&& a^- a^+ = E(N+1) - E_0, \\
&& [a^-,a^+] = E(N+1) - E(N) \equiv f(N),  \label{commapam} \\
&& [H,a^\pm] = \pm f(N - 1/2 \mp 1/2) a^\pm.
\end{eqnarray}

Let us note that, depending of the key function $E(n)$ associated to the initial Hamiltonian, the system 
could be ruled by a Lie algebra, in case that $E(n)$ is either linear or quadratic in $n$. However, it 
could be also ruled by non-Lie algebras, when $E(n)$ has a more involved dependence with $n$.

Once we have characterized the algebra for the initial Hamiltonian, it is possible to analyze the 
corresponding structure for its SUSY partner Hamiltonians $\widetilde H$.

\subsection{Algebraic structure of $\widetilde H$}

The most important properties of $\widetilde H$ come from its connection with the initial 
Hamiltonian $H$ through the intertwining operators (see equation~(\ref{intertwining})). In fact, from 
these expressions it is simple to identify the {\it natural} ladder operators for $\widetilde H$ as 
follows \cite{mi84,fhn94,fh99,fhr07}:
\begin{eqnarray}
&& {\widetilde a}^\pm = B^+ a^\pm B. \label{natladop}
\end{eqnarray}
Its action on the eigenstates of $\widetilde H$ can be straightforwardly calculated, leading to: 
\begin{eqnarray}
& {\widetilde a}^\pm \vert \widetilde\psi_{\epsilon_j}\rangle = 0, \\
&  {\widetilde a}^- \vert \widetilde\psi_n \rangle = {\widetilde r}(n) \vert \widetilde\psi_{n-1} \rangle, \\
&  {\widetilde a}^+ \vert \widetilde\psi_n\rangle = {\widetilde r}^{\,*}(n + 1) \ \vert \widetilde\psi_{n+1}\rangle,  \\
& {\widetilde r}(n) = \left[\prod\limits_{i=1}^k[E(n) -\epsilon_i][E(n-1)-\epsilon_i]\right]^{\frac12} \, r(n). 
\label{tilder}
\end{eqnarray}
In order to simplify the discussion, from now on we will assume that none of the $\epsilon_j, \,
j=1,\dots,k$ coincide with some eigenvalue of $H$, and that $k$ new energy levels are created for 
$\widetilde H$ at $\epsilon_j, j=1,\dots,k$. It is important as well to define the number operator 
$\widetilde N$ for the system ruled by $\widetilde H$, through its action on the corresponding energy 
eigenstates:
\begin{eqnarray}
&& {\widetilde N}\vert\widetilde\psi_{\epsilon_j}\rangle = 0, \\
&& {\widetilde N}\vert\widetilde\psi_n\rangle = n \vert\widetilde\psi_n\rangle.
\end{eqnarray}

The {\it natural algebra} of the system is now defined by:
\begin{eqnarray}
& [{\widetilde N},{\widetilde a}^\pm] = \pm  {\widetilde a}^\pm, \\
& [{\widetilde a}^-, {\widetilde a}^+] = \left[{\widetilde r}^{\,*}({\widetilde N}+1) \, {\widetilde r}({\widetilde N}+1) 
- {\widetilde r}^{\,*}({\widetilde N}) \, {\widetilde r}({\widetilde N})\right] 
\sum\limits_{n=0}^{\infty}\vert\widetilde\psi_n\rangle \, \langle\widetilde\psi_n\vert , \label{complepart}
\end{eqnarray}
where ${\widetilde r}(n)$ is given by equations~(\ref{tilder},\ref{phfacal}).

\subsection{Coherent states of $H$ and $\widetilde H$}

We have just identified the annihilation and creation operators for the SUSY partner Hamiltonians $H$ 
and $\widetilde H$. The coherent states for such systems can be looked for as eigenstates of the 
annihilation operator with complex eigenvalues $z$, namely:
\begin{eqnarray}
& a^- \vert z,\tau\rangle = z \vert z,\tau\rangle, \label{bgcs-ini} \\
& {\widetilde a}^- \vert \widetilde{z,\tau}\rangle = z \vert \widetilde{z,\tau}\rangle. \label{bgcs-fin}
\end{eqnarray}
If we expand the coherent states in the basis of energy eigenstates, substitute them in 
equations~(\ref{bgcs-ini},\ref{bgcs-fin}) to obtain a recurrence relation for the coefficients of the 
expansion, express such coefficients in terms of the first one and normalize them, we arrive at the 
following expressions:
\begin{eqnarray}
\vert z,\tau\rangle & = &
\left(\sum\limits_{m=0}^{\infty} \frac{\vert
z\vert^{2m}}{\rho_m}\right)^{-\frac12} \sum\limits_{m =
0}^{\infty}e^{-i\tau (E_m - E_0)}
\frac{z^m}{\sqrt{\rho_m}}\vert\psi_m\rangle , \label{csini} \\
\rho_m & = & \begin{cases} 1 & {\rm if} \ m=0 \cr (E_m - E_0)\cdots(E_1 -
E_0) & {\rm if} \ m > 0 \end{cases}
\end{eqnarray}
and
\begin{eqnarray}
\vert \widetilde{z,\tau}\rangle & = & \left(\sum\limits_{m = 0}^{\infty} \frac{\vert
z\vert^{2m}}{\widetilde\rho_m}\right)^{-\frac12} \sum\limits_{m
= 0}^{\infty} e^{-i\tau (E_{m} -
E_{0})}\frac{z^{m}}{\sqrt{\widetilde\rho_m}}\vert\widetilde\psi_{m}\rangle ,  \label{csfin} \\
\widetilde\rho_m  & = & 
\begin{cases} 1 & {\rm if} \ m=0 \cr
\rho_{m} \prod\limits_{i=1}^k (E_{m} - \epsilon_i) (E_{m - 1} - \epsilon_i)^2 \dots (E_{1} - \epsilon_i)^2 
(E_{0} - \epsilon_i) & {\rm if} \ m>0 \end{cases}
\end{eqnarray}

It is important to ensure that our coherent states fulfill a completeness relation, in order that an 
arbitrary state can be decomposed in terms of them. In our case the two completeness relations are:
\begin{eqnarray}
&& \int \vert z,\tau\rangle \langle z,\tau \vert d\mu(z) = 1 , \\
&& d\mu(z) = \frac{1}{\pi} \left(\sum\limits_{m=0}^{\infty}\frac{\vert
z\vert^{2m}}{\rho_m}\right) \rho(\vert z\vert^2) \, d^2 z,
\end{eqnarray}
and
\begin{eqnarray}
&& \sum\limits_{i=1}^k \vert \widetilde\psi_{\epsilon_i}\rangle \langle
\widetilde\psi_{\epsilon_i} \vert + \int \vert \widetilde{z,\tau}\rangle
\langle \widetilde{z,\tau} \vert \, d\widetilde\mu(z) = 1 , \\
&& d\widetilde\mu(z) = \frac{1}{\pi} \left(\sum\limits_{m =
0}^{\infty} \frac{\vert z\vert^{2m}}{\widetilde\rho_m}\right)
\widetilde\rho(\vert z\vert^2) \, d^2z .
\end{eqnarray}
They will be fulfilled if we would find two measure functions $\rho(y)$ and $\widetilde\rho(y)$ solving 
the following moment problems \cite{fhn94,fh99,sps99,sp00,kps01}
\begin{eqnarray}
&& \int_0^\infty y^m \rho(y) \, dy = \rho_m, \label{momentsH} \\
&& \int_0^\infty y^m \widetilde\rho(y) \, dy = \widetilde \rho_m, 
\qquad m = 0,1,\dots \label{momentstildeH}
\end{eqnarray}
The fact that two coherent states of a given family in general are not orthogonal is contained in the 
so-called reproducing kernel, which turns out to be:
\begin{eqnarray}
\langle z_1,\tau \vert z_2,\tau\rangle & = &
\left(\sum\limits_{m=0}^{\infty} \frac{\vert z_1
\vert^{2m}}{\rho_m}\right)^{-\frac12}
\left(\sum\limits_{m=0}^{\infty} \frac{\vert
z_2\vert^{2m}}{\rho_m}\right)^{-\frac12}
\left(\sum\limits_{m=0}^{\infty} \frac{(\bar z_1z_2)^m}{\rho_m}\right), \\
\langle \widetilde{z_1,\tau} \vert \widetilde{z_2,\tau}\rangle & = &
\left(\sum\limits_{m=0}^{\infty} \frac{\vert z_1\vert^{2m}}{\widetilde\rho_m}\right)^{-\frac12}
\left(\sum\limits_{m=0}^{\infty} \frac{\vert z_2\vert^{2m}}{\widetilde\rho_m}\right)^{-\frac12}
\left(\sum\limits_{m=0}^{\infty} \frac{(\bar z_1z_2)^m}{\widetilde\rho_m}\right).
\end{eqnarray}
Concerning dynamics, the coherent states evolve as follows:
\begin{eqnarray}
U(t) \vert z,\tau\rangle & = & \exp(-itH) \vert z,\tau\rangle = e^{-itE_0}\vert z,\tau+t\rangle, \\
\widetilde U(t) \vert \widetilde{z,\tau}\rangle & = & \exp(-it\widetilde H) \vert \widetilde{z,\tau}\rangle = 
e^{-itE_0}\vert\widetilde{z,\tau + t}\rangle.
\end{eqnarray}

Let us note that, while the eigenvalue $z=0$ of $a^-$ is non-degenerate (if $z=0$ is made in 
equation~(\ref{csini}) the ground state of $H$ is achieved), for $\widetilde a^-$ this eigenvalue is 
$(k+1)$th degenerate, since all states $\widetilde\psi_{\epsilon_i}, i=1,\dots,k$ are annihilated by 
$\widetilde a^-$ and for $z=0$ equation~(\ref{csfin}) reduces to the eigenstate 
$|\widetilde\psi_0\rangle$ of $\widetilde H$ associated to $E_0$.

\subsection{Example: harmonic oscillator}
The simplest system available to illustrate the previous treatment is the harmonic oscillator. In this 
case there is a linear relation between the number operator and the Hamiltonian $H$, $H = E(N) = N+1/2$. 
In addition, the function characterizing the action of $a^\pm$ onto the eigenstates of $H$ becomes:
\begin{eqnarray}\label{hopot}
&& r(n) = \sqrt{E_n - E_0} = \sqrt{n},
\end{eqnarray}
where, since the phase factors of equation~(\ref{phfacal}) are independent of $n$, we have fixed them 
by taking $\tau=0$. The function characterizing the commutator between the annihilation and creation 
operators is now (see equation~(\ref{commapam})):
\begin{eqnarray}
&& f(N) = E(N + 1) - E(N) = 1. \label{fnpt}
\end{eqnarray}
Thus, the commutation relations for the intrinsic algebra of the oscillator becomes: 
\begin{eqnarray}
& [N, a^\pm] =  \pm a^\pm, \\ 
& [a^-, a^+] = 1,
\end{eqnarray}
which is the well known Heisenberg-Weyl algebra.

On the other hand, for the SUSY partner Hamiltonian $\widetilde H$ we have that:
\begin{eqnarray}
& \widetilde r(n) = \left[\prod\limits_{i=1}^k \left(E_n - \epsilon_i - 1\right)\left(E_n
- \epsilon_i  \right)\right]^{\frac12} r(n).
\end{eqnarray}
If we insert this expression in equation~(\ref{complepart}) it is obtained a polynomial Heisenberg 
algebra, since in this case the commutator of $\widetilde a^-$ and $\widetilde a^+$ is a polynomial of 
degree $2k$ either in $\widetilde H$ or in $\widetilde N$.

Concerning coherent states, in the first place the coefficients $\rho_m$ and $\widetilde\rho_m$, 
which are also the moments arising in equations~(\ref{momentsH},\ref{momentstildeH}), become:
\begin{eqnarray}
& \rho_m = m!, \\
& \widetilde\rho_m  = m! \prod\limits_{i=1}^k(\frac12 - \epsilon_i)_m (\frac32 - \epsilon_i)_m ,
\end{eqnarray}
where $(c)_m=\Gamma(c+m)/\Gamma(c)$ is a Pochhammer's symbol. It is straightforward to find now the 
explicit expressions for the coherent states:
\begin{eqnarray}
&& \vert z\rangle = e^{-\frac{|z|^2}{2}}\sum\limits_{m = 0}^\infty 
\frac{z^m}{\sqrt{m!}}\vert\psi_m\rangle, \\
&& {\vert \widetilde z\rangle} = \sum\limits_{m = 0}^{\infty} 
\frac{z^m\vert\widetilde\psi_{m}\rangle}{\sqrt{{}_0F_{2k}(\frac12-\epsilon_1, \frac32 -\epsilon_1,\dots,
\frac12-\epsilon_k,\frac32-\epsilon_k;\vert z \vert^2) \, m! \prod\limits_{i=1}^k 
(\frac12-\epsilon_i)_m(\frac32-\epsilon_i)_m}}.
\end{eqnarray}
The solutions to the moment problems of equations~(\ref{momentsH},\ref{momentstildeH}) are given by:
\begin{eqnarray}
&& \rho (y) = \exp\left(-y\right) , \\
&& \widetilde\rho (y) = \frac{ G^{2k+1 \ \ 0}_{\ \ 0 \ \ 2k+1}(y\vert
0,-\epsilon_1  - \frac12,\dots,-\epsilon_k - \frac12,\frac12 - \epsilon_1,\dots,\frac12 - \epsilon_k)} 
{\prod\limits_{i=1}^k\Gamma(\frac12-\epsilon_i)\Gamma(\frac32-\epsilon_i)} ,
\end{eqnarray}
where $G$ is a Meijer G-function \cite{fh99}. The reproducing kernel in both cases turn out to be:
\begin{eqnarray}
& \langle z_1 \vert z_2 \rangle = \exp\left[-\frac12(|z_1|^2 + |z_2|^2- 2 z_1^{\,*}z_2)\right], \\
& \langle \widetilde{z_1}\vert \widetilde{z_2}\rangle  = \frac{{}_0F_{2k}(\frac12-\epsilon_1, \frac32 -\epsilon_1,\dots,
\frac12-\epsilon_k,\frac32-\epsilon_k;z_1^{\,*}z_2)}
{\sqrt{{}_0F_{2k}(\frac12-\epsilon_1, \frac32 -\epsilon_1,\dots,
\frac12-\epsilon_k,\frac32-\epsilon_k;|z_1|^2) 
{}_0F_{2k}(\frac12-\epsilon_1, \frac32 -\epsilon_1,\dots,
\frac12-\epsilon_k,\frac32-\epsilon_k; |z_2|^2)}}  .
\end{eqnarray}

\begin{figure}[ht]
\centering 
\epsfig{file=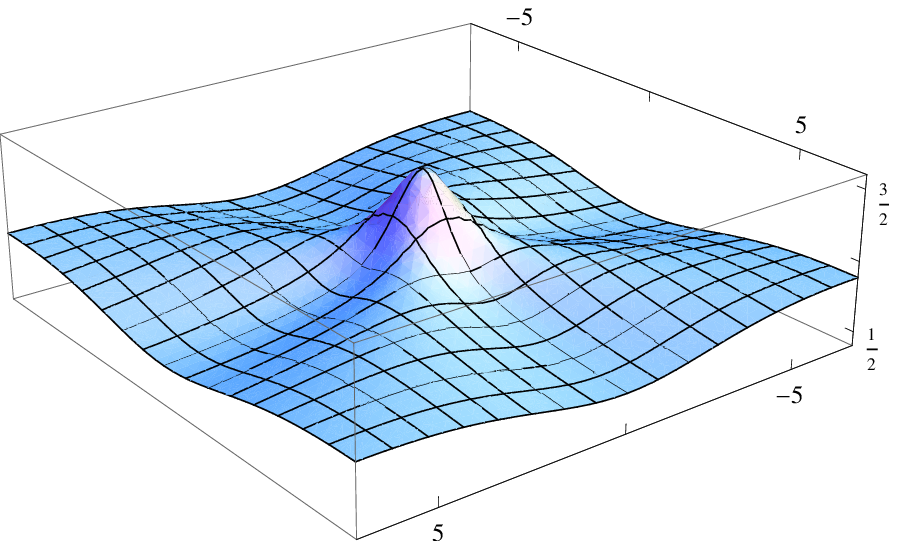, width=10cm}
\caption{Uncertainty relation $(\Delta X)(\Delta P)$ for the coherent states ${\vert \widetilde 
z\rangle}$ with $k=1$ in the harmonic oscillator limit, when $\epsilon_1=-\frac12, 
\nu_1=0$.}
\label{1csdxdpk}
\end{figure}

\begin{figure}[ht]
\centering 
\epsfig{file=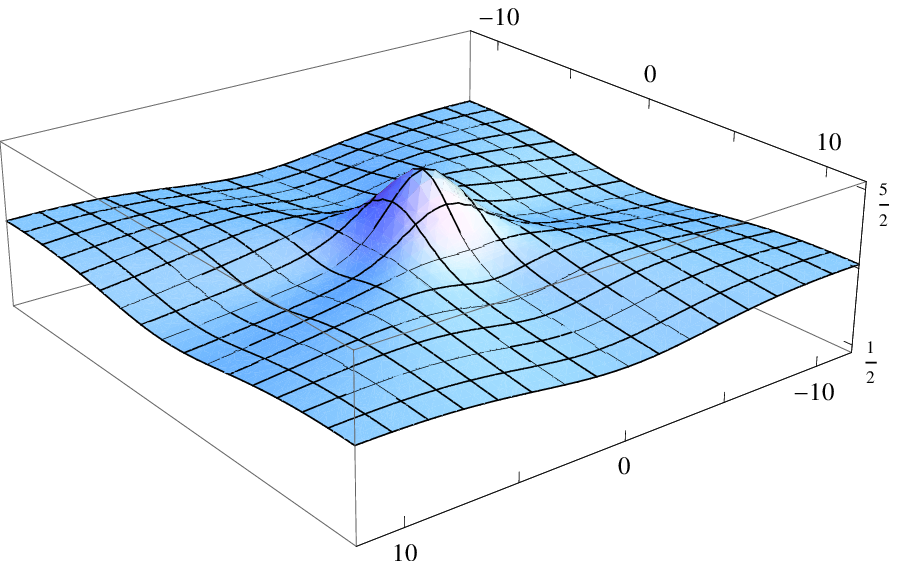, width=10cm}
\caption{Uncertainty relation $(\Delta X)(\Delta P)$ for the coherent states ${\vert \widetilde 
z\rangle}$ with $k=2$ in the harmonic oscillator limit, when $(\epsilon_1,\epsilon_2) = (-\frac12,-\frac32), \ 
(\nu_1,\nu_2)= (0,\infty)$.}
\label{2csdxdpk}
\end{figure}

As we can see, the coherent states for the initial Hamiltonian $H$ are the standard ones, which minimize 
the Heisenberg uncertainty relation, i.e., $(\Delta X) (\Delta P)=1/2$. It would be important to know if 
the coherent states associated to $\widetilde H$ have also this property. However, the calculation of 
$(\Delta X) (\Delta P)$ for general SUSY transformations, with arbitrary factorization energies and 
associated constants $\epsilon_j, \, \nu_j, j=1,\dots,k$ involved in the Schr\"odinger solution of 
equation~(\ref{gsseho}), is difficult. Despite, such an uncertainty can be analytically calculated in 
the harmonic oscillator limit for an arbitrary $k$. In particular, for $k=1$, $\epsilon_1=-\frac12, 
\nu_1=0$ it is obtained \cite{fhn94} ($r=|z|$):
\begin{eqnarray}
&(\Delta X)(\Delta P) = \sqrt{\{\frac32 - [{\rm Re}(z)]^2\xi_1(r)\}\{\frac32 - [{\rm Im}(z)]^2\xi_1(r)\}},  
\label{DxDp1}\\
& \xi_1(r) = 2 \left[
\frac{{}_0F_2(2,2;r^2)}{{}_0F_2(1,2;r^2)}\right]^2 -
\left[\frac{{}_0F_2(2,3;r^2)}{{}_0F_2(1,2;r^2)}\right] ,
\end{eqnarray}
while for $k=2$, $(\epsilon_1,\epsilon_2) = (-\frac12,-\frac32), \ (\nu_1,\nu_2)= (0,\infty)$ we arrive 
at \cite{fh99}:
\begin{eqnarray}
&(\Delta X)(\Delta P) = \sqrt{\{\frac52 - [{\rm Re}(z)]^2\xi_2(r)\}\{\frac52 - [{\rm Im}(z)]^2\xi_2(r)\}},  
\label{DxDp2}\\
& \xi_2(r) = \frac12 \left[
\frac{{}_0F_4(2,2,3,3;r^2)}{{}_0F_4(1,2,2,3;r^2)}\right]^2 -
\frac16 \left[\frac{{}_0F_4(2,3,3,4;r^2)}{{}_0F_4(1,2,2,3;r^2)}\right] .
\end{eqnarray}
Plots of the Heisenberg uncertainty relations of equations~(\ref{DxDp1}) and (\ref{DxDp2}) as functions 
of $z$ are shown in figures \ref{1csdxdpk} and \ref{2csdxdpk} respectively. It is seen that these 
coherent states are no longer minimum uncertainty states. However, for $k=1$ there are some directions 
in the complex plane for which the minimum value $(\Delta X) (\Delta P)=1/2$ is achieved when 
$|z|\rightarrow\infty$ (see Fig.~1). 

\section{SUSY QM and Painlev\'e equations}

In a general context, the polynomial Heisenberg algebras (PHA) of degree $m$ are deformations of the 
Heisenberg-Weyl algebra for which the commutators of the Hamiltonian $H$ (of form given in 
equation~(\ref{HandHt})) with $(m+1)$th order differential ladder operators $L^\pm$ are standard 
while the commutator between $L^-$ and $L^+$ is a polynomial of degree $m$th in $H$ \cite{cfnn04}, i.e.,
\begin{eqnarray}
&& [H,L^\pm] = \pm L^\pm,  \label{phagral1}\\
&& [L^-,L^+] = q_{m+1}(H+1)-q_{m+1}(H)= p_m(H), \label{phagral2}\\
&& L^+ L^- = q_{m+1}(H) = \prod\limits_{j=1}^{m+1} (H-{\cal E}_j), \label{phagral3}\\
&& L^- L^+ = q_{m+1}(H+1) = \prod\limits_{j=1}^{m+1} (H-{\cal E}_j+1). \label{phagral4}
\end{eqnarray}
Systems ruled by PHA of degree $m$ have $m+1$ extremal states $\psi_{{\cal E}_j}, \, j=1,\dots,m+1$, 
which are annihilated by $L^-$ and are formal eigenstates of $H$ associated to ${\cal E}_j$.

Previously it was shown that the SUSY partner Hamiltonians of the harmonic oscillator are ruled by PHA 
of degree $2k$, with their natural ladder operators being of order $2k+1$ (see 
equation~(\ref{natladop})). Hence, the first order SUSY partners of the harmonic oscillator are ruled by 
second-degree polynomial Heisenberg algebras generated by third-order ladder operators, and so on. Thus, 
through SUSY QM plenty of particular realizations of such algebras can be supplied. However, it would be 
important to identify the general Hamiltonians $H$, of form given in equation~(\ref{HandHt}), which have 
$(m+1)$th order differential ladder operators. This question has been addressed recurrently in the past, 
and nowadays there are some definite answers: if $m=0$ the general potential having first-order ladder 
operators is the harmonic oscillator, while for $m=1$ (second-order ladder operators) it is the radial 
oscillator. On the other hand, for $m=2$ ($m=3$) the general potential with third-order (fourth-order) 
ladder operators is expressed in terms of a function which satisfies the Painlev\'e IV (V) equation 
\cite{cfnn04}. 

This connection suggests the possibility of going in the inverse direction, so if we could identify a 
Hamiltonian with third-order (fourth-order) ladder operators, perhaps we could use some information (the 
extremal state expressions and associated factorization energies ${\cal E}_j$) to generate solutions to 
the Painlev\'e IV (V) equation (also called Painlev\'e IV (V) transcendents). This is in fact what 
happens, thus the game reduces to find Hamiltonians with third-order (fourth-order) ladder operators for 
generating Painlev\'e IV (V) transcendents through the extremal states of the system 
\cite{cfnn04,bf11b,bf14,bfn16}. 

Let us present next these statements as two algorithms to generate solutions for such non-linear 
second-order ordinary differential equations.

\subsection{Generation of Painlev\'e IV transcendents}

Let us suppose that we have identified a Hamiltonian of the form given in equation~(\ref{HandHt}), which 
has third-order differential ladder operators $L^\pm$ satisfying 
equations~(\ref{phagral1}-\ref{phagral4}) with $m=2$, as well as its three extremal states 
$\psi_{{\cal E}_j}$ and associated factorization energies ${\cal E}_j, \, j=1,2,3$. Thus, a solution to 
the Painlev\'e IV (PIV) equation
\begin{eqnarray}
&& g''=\frac{{g'}^2}{2g} + \frac32 g^3 + 4 x g^2 + 2(x^2-\alpha) g + \frac{\beta}{g}, 
\end{eqnarray}
is given by
\begin{eqnarray}
&& g(x) = - x -\{\ln[\psi_{{\cal E}_3}(x)]\}', 
\end{eqnarray}
where the parameters $\alpha,\beta$ of the PIV equation are related with ${\cal E}_1, \, {\cal E}_2, \, 
{\cal E}_3$ 
in the way
\begin{eqnarray}
&& \alpha={\cal E}_1 + {\cal E}_2 - 2 {\cal E}_3 -1, \qquad \beta= -2({\cal E}_1 - {\cal E}_2)^2. 
\end{eqnarray}
Let us note that, if the indices asigned to the extremal states are permuted cyclically, we will obtain 
three PIV transcendents, one for each extremal state when it is labeled as $\psi_{{\cal E}_3}$.

Summarizing, our task has been reduced to identify systems ruled by second degree PHA and the 
corresponding extremal states \cite{cfnn04,bf11b}. The harmonic oscillator supplies several such possibilities, for 
instance, the two operator pairs $\{a^3,(a^+)^3\}$, $\{a^+a^2,(a^+)^2a\}$ are third order ladder 
operators satisfying equations~(\ref{phagral1}-\ref{phagral4}) (the level spacing has to be adjusted in 
the first case), and it is simple to identify the corresponding extremal states. On the other hand, the 
first-order SUSY partners of the oscillator also have natural third-order ladder operators, and well 
identified extremal states. For the SUSY partners of the oscillator with $k\geq 2$ the natural ladder 
operators are not of third order (they are in general of order $2k+1$). However, it is possible to 
induce a reduction process, by choosing connected seed solutions $u_{j+1}= au_j, \ \epsilon_{j+1} = 
\epsilon_{j}-1, \ j=1,\dots,k-1$ instead of general seed solutions, so that the $(2k+1)$th order 
ladder operators reduce to third order ones.

Some examples of real PIV transcendents associated to real PIV parameters $\alpha,\ \beta$, which are 
generated through this algorithm, are presented next.

\subsubsection{Harmonic oscillator.} If we take the ladder operators $L^-=a^3,L^+=(a^+)^3$ for the 
harmonic oscillator Hamiltonian we get the PIV transcendents reported in Table~1 \cite{cf17}. Note that 
in order that the level spacing induced by this pair of ladder operators coincide with the standard one 
($\Delta E=1$) of equations~(\ref{phagral1}-\ref{phagral4}), we need to change variables $y=\sqrt{3}x$ 
and scale the factorization energies (dividing by $3$). Remember also that $\psi_j(x)$ are the 
eigenfunctions of the harmonic oscillator associated to the first three energy levels $E_j=j+1/2, 
j=0,1,2$.

\begin{table}[h] 
\centering
\begin{tabular}{cccc}
\hline \\[-12pt]
$\psi_{{\cal E}_3}$ & $\psi_0(x)$ & $\psi_1(x)$ & $\psi_2(x)$ \\ [3pt]
${\cal E}_3$ & $\frac12$ &  $\frac32$ &  $\frac52$ \\ [3pt]
$g(y)$ & $-\frac{2y}3$ &  $-\frac{2y}3 - \frac{1}{y}$ &  $-\frac{2y}3 - \frac{4y}{2y^2-3}$ \\ [3pt]
$\alpha$ & $0$ & $-1$ & $-2$ \\ [3pt]
$\beta$ & $-\frac29$ & $-\frac89$ & $-\frac29$ \\ [3pt]
\hline
\end{tabular}
\caption{PIV transcendents generated from the harmonic oscillator Hamiltonian with 
$L^-=a^3,L^+=(a^+)^3$.}
\end{table}

\subsubsection{First-order SUSY partner of the harmonic oscillator.} For $\epsilon_1=-\frac52, \ 
\nu_1=0$ and the third-order ladder operators $L^-=B^+aB, \ L^+=B^+a^+B$ of $\widetilde H$, we get the 
PIV transcendents reported in Table~2. The seed solution employed is 
$u_1(x) = e^{\frac{x^2}2} (1+2 x^2)$.

\begin{table}[h] 
\centering
\begin{tabular}{cccc}
\hline \\[-12pt]
$\psi_{{\cal E}_3}$  & $\frac{1}{u_1}$             & $B^+\psi_0$                  & $B^+a^+u_1$ \\ [3pt]
${\cal E}_3$         & $-\frac52$          &  $\frac12$                   & $-\frac32$  \\ [3pt]
$g(x)$ & $\frac{4x}{1+2x^2}$ & $-\frac{4 x^4+3}{4 x^5+8 x^3+3 x}$ & $\frac{8 x^5+6 x}{1-4 x^4}$ \\ [3pt]
$\alpha$                  & $3$                 & $-6$                         & $0$ \\ [3pt]
$\beta$                  & $-8$                & $-2$                         & $-18$ \\ [3pt]
\hline
\end{tabular}
\caption{PIV transcendents generated from the first-order SUSY partner Hamiltonian $\widetilde H$ with 
$L^-=B^+aB, \ L^+=B^+a^+B$.}
\end{table}

\subsubsection{Second-order SUSY partner of the harmonic oscillator.} For $\epsilon_1=-\frac52, \ 
\nu_1=0$ and the third-order ladder operators of $\widetilde H$ obtained from the reduction of the 
fifth-order ones $L^-=B^+aB, \ L^+=B^+a^+B$, we get the PIV transcendents reported in Table~3. Once 
again, the seed solution $u_1$ employed is $u_1(x) = e^{\frac{x^2}2} (1+2 x^2)$ and $u_2= a u_1$.

\begin{table}[h] 
\centering
\begin{tabular}{cccc}
\hline \\[-12pt]
$\psi_{{\cal E}_3}$ & $\frac{u_1}{W(u_1,u_2)}$    & $B^+\psi_0$                  & $B^+a^+u_1$ \\ [3pt]
${\cal E}_3$        & $-\frac72$          &  $\frac12$                   & -$\frac32$  \\ [3pt]
$g(x)$ & $\frac{4 x \left(4 x^4+4 x^2-3\right)}{8 x^6+4 x^4+6 x^2+3}$ & $-\frac{4 x \left(16 x^8+72 x^2+27\right)}{32 x^{10}+48 x^8+96 x^6+54 x^2-27}$ & $\frac{-16 x^8+32 x^6-48 x^4+9}{x \left(2 x^2-3\right) \left(4 x^4+3\right)}$ \\ [3pt]
$\alpha$                 & $5$                 & $-7$                         & $-1$ \\ [3pt]
$\beta$                 & $-8$                & $-8$                         & $-32$ \\ [3pt]
\hline
\end{tabular}
\caption{PIV transcendents generated from the second-order SUSY partner Hamiltonian $\widetilde H$ and 
the third-order ladder operators obtained by reducing $L^-=B^+aB, \ L^+=B^+a^+B$.}
\end{table}

\subsection{Generation of Painlev\'e V transcendents}

Let us suppose now that the Hamiltonian $H$ we have identified has fourth-order ladder operators and 
satisfy equations~(\ref{phagral1}-\ref{phagral4}) with $m=3$. We know also its four extremal states 
$\psi_{{\cal E}_j}$ and associated factorization energies ${\cal E}_j, \, j=1,2,3,4$. Thus, one 
solution to the Painlev\'e V (PV) equation 
\begin{eqnarray}\label{PVgral}
&& w''=\left(\frac{1}{2w}-\frac{1}{w-1}\right)(w')^2 - \frac{w'}{z} + \frac{(w-1)^2}{z^2} 
\left(\alpha \, w+\frac{\beta}{w}\right) + \gamma \frac{w}{z} + \delta \frac{w(w+1)}{w-1},
\end{eqnarray}
is given by
\begin{eqnarray}
w(z) & = & 1 + \frac{\sqrt{z}}{g(\sqrt{z})}, \\
g(x) & = & - x -\frac{d}{dx}\left\{\ln\left[W(\psi_{{\cal E}_3}(x),\psi_{{\cal E}_4}(x))\right]\right\}, 
\end{eqnarray}
where the prime in equation~(\ref{PVgral}) means derivative with respect to $z$, and the PV parameters 
$\alpha, \ \beta, \ \gamma, \ \delta$ are related with ${\cal E}_1, \, {\cal E}_2, \, {\cal E}_3, \, 
{\cal E}_4$ through
\begin{eqnarray}
&& \alpha=\frac{({\cal E}_1 - {\cal E}_2)^2}2, \quad  \beta = -\frac{({\cal E}_3 - {\cal E}_4)^2}2, \quad  
\gamma = \frac{{\cal E}_1 + {\cal E}_2}2 - \frac{{\cal E}_3 + {\cal E}_4 + 1}2, \quad \delta = - \frac18. 
\end{eqnarray}
Note that if the indices of the extremal states are permuted, we will obtain at the end six PV 
transcendents (in principle different), one for each pair of extremal states when they are labeled as 
$\psi_{{\cal E}_3}, \ \psi_{{\cal E}_4}$ \cite{bfn16}.

Once again, now we require just to identify systems ruled by third degree PHA and their four extremal 
states. The harmonic oscillator also supplies some possibilities, the simplest one through the fourth 
order ladder operators $\{L^-=a^4,L^+=(a^+)^4\}$, which satisfy 
equations~(\ref{phagral1}-\ref{phagral4}) if we change variables and adjust the levels spacing, with the 
extremal states being the eigenstates associated to the four lowest energy levels of the oscillator. 
Another system closely related to PV equation is the radial oscillator, for which its ladder operators 
$b^\pm$ are of second order \cite{bfn16}. Thus, the second powers of such operators are also fourth 
order ladder operators that will give place to PV transcendents. Concerning SUSY partners, those of the 
radial oscillator give place to PHA of degree $2k+1$, with natural ladder operators of order $2k+2$. 
Thus, the first order SUSY partners of the radial oscillator have natural fourth order ladder operators 
and well identified extremal states. For $k\geq 2$, it is possible to produce again a reduction process, 
by connecting the seed solutions in the way $u_{j+1}= b^-u_j, \ \epsilon_{j+1} = \epsilon_{j}-1, \ 
j=1,\dots,k-1$, so that the $(2k+2)$th order natural ladder operators reduce to fourth order ones 
\cite{bfn16}. Remember that the first-order SUSY partners of the harmonic oscillator also have 
fourth-order ladder operators, given by $L^-=B^+ a^2 B, \ L^+=B^+ (a^+)^2 B$, but we will have to change 
variables and adjust the level spacing to stick to the standard convention $\Delta E=1$.

Some examples of real PV transcendents associated to real parameters $\alpha, \ \beta, \ \gamma, \ \delta$, 
generated through this algorithm, are now presented.

\subsubsection{Harmonic oscillator.} If we take $L^-=a^4,L^+=(a^+)^4$ as ladder operators, we generate 
the PV transcendents reported in Table~4. Note that here $z=4x^2$ and $\psi_j(x), j=0,1,2,3$ are the 
eigenfunctions for the four lowest eigenvalues of the harmonic oscillator. We initially order the extremal 
states as
\begin{eqnarray}
& \psi_{{\cal E}_1}(x) = \psi_2(x), \qquad {\cal E}_1 = \frac52, \\
& \psi_{{\cal E}_2}(x) = \psi_3(x), \qquad {\cal E}_2 = \frac72, \\ 
& \psi_{{\cal E}_3}(x) = \psi_0(x), \qquad {\cal E}_3 = \frac12, \\
& \psi_{{\cal E}_4}(x) = \psi_1(x), \qquad {\cal E}_4 = \frac32,
\end{eqnarray}
and this permutation will be denoted as $1234$. We do not include the parameter $\delta$ in this table since 
it is constant ($\delta=-\frac18$).

\begin{table}[h] 
\centering
\begin{tabular}{ccccc}
\hline \\[-12pt]
Permutation &  $\alpha$  & $\beta$  & $\gamma$  & $w(z)$ \\
\hline \\[-12pt]
$1234$ & $\frac{1}{32}$ & $-\frac{1}{32}$ & $0$ & $-1$ \\ [3pt]
$4231$ & $\frac18$ &  $-\frac18$ &  $-\frac14$ & $\frac{2-z}{z+2}$ \\ [3pt]
$1432$ & $\frac{1}{32}$ &  $-\frac{9}{32}$ &  $-\frac12$ &  $\frac{6-z}{z+2}$ \\ [3pt]
$3241$ & $\frac{9}{32}$ &  $-\frac{1}{32}$ &  $-\frac12$ &  $\frac{2-z}{z+6}$ \\ [3pt]
$3142$ & $\frac{1}{8}$ &  $-\frac{1}{8}$ &  $-\frac34$ &  $\frac{6-z}{z+6}$ \\ [3pt]
$3412$ & $\frac{1}{32}$ &  $-\frac{1}{32}$ &  $-1$ &  $-\frac{(z-6) (z-2)}{(z+2) (z+6)}$ \\ [3pt]
\hline
\end{tabular}
\caption{PV transcendents generated from the harmonic oscillator Hamiltonian and $L^-=a^4,L^+=(a^+)^4$.}
\end{table}

\subsubsection{First-order SUSY partner of the harmonic oscillator.} For $\epsilon_1=-\frac52, \ 
\nu_1=0$ and the fourth-order ladder operators $L^-=B^+a^2B,\ L^+=B^+(a^+)^2B$ of $\widetilde H$, we 
will get the PV transcendents reported in Table~5, where $z=2x^2$. The seed solution employed is $u_1(x)= 
e^{\frac{x^2}2} (1+2 x^2)$. The initial order for the extremal states, denoted as $1234$ in the table, 
is
\begin{eqnarray}
\psi_{{\cal E}_1}(x) & = & \frac{W(u_1,\psi_0)}{u_1}, \qquad {\cal E}_1 = \frac12, \\
\psi_{{\cal E}_2}(x) & = & \frac{W(u_1,\psi_1)}{u_1}, \qquad {\cal E}_2 = \frac32, \\ 
\psi_{{\cal E}_3}(x) & = & \frac{1}{u_1}, \hskip2.2cm {\cal E}_3 = -\frac52, \\
\psi_{{\cal E}_4}(x) & = & B^+ (a^+)^2 u_1, \qquad {\cal E}_4 = -\frac12.
\end{eqnarray}

\begin{table}[h] 
\centering
\begin{tabular}{ccccc}
\hline \\[-12pt]
Permutation &  $\alpha$  & $\beta$  & $\gamma$  & $w(z)$ \\
\hline \\[-12pt]
$1234$ & $\frac{1}{8}$ & $-\frac{1}{2}$ & $\frac34$ & $-\frac{2}{z-1}$ \\ [3pt]
$4231$ & $\frac12$ &  $-\frac98$ &  $\frac14$ & $\frac{z+3}{2}$ \\ [3pt]
$1432$ & $\frac{1}{8}$ &  $-2$ &  $-\frac14$ &  $\frac{z^2+2 z-1}{z-1}$ \\ [3pt]
$3241$ & $2$ &  $-\frac{1}{8}$ &  $-\frac34$ &  $\frac{z+3}{z^2+2 z+3}$ \\ [3pt]
$3142$ & $\frac{9}{8}$ &  $-\frac{1}{2}$ &  $-\frac54$ &  $\frac{2 \left(z^2+2 z-1\right)}{z^3+z^2+z-3}$ \\ [3pt]
$3412$ & $\frac{1}{2}$ &  $-\frac{1}{8}$ &  $-\frac74$ &  $-\frac{z^3+5 z^2+5 z-3}{2 \left(z^2+2 z+3\right)}$ \\ [3pt]
\hline
\end{tabular}
\caption{PV transcendents generated from the first-order SUSY partner Hamiltonian $\widetilde H$ of the 
oscillator and $L^-=B^+a^2B,\ L^+=B^+(a^+)^2B$.}
\end{table}

We conclude this section by stating that an infinity of PIV and PV transcendents can be derived through 
the techniques described here. It is an open question to determine if any exact solution to such 
equations that exists in the literature can be derived through these methods. However, the algorithms 
are so simple and direct that we felt it was the right time to try to make them known to a wider and 
diversified community, not just to people working on solutions to non-linear differential equations.

\section{Recent applications of SUSY QM}

Some recent interesting applications of SUSY QM are worth of some discussion. We would like to mention 
in the first place the motion of electrons in graphene, a single layer of carbon atoms arranged in a 
hexagonal honeycomb lattice. Since close to the Dirac points in the Brillouin zone there is a gapless linear 
dispersion relation, obtained in the low energy regime through a tight binding model, one ends up with an 
electron description in terms of the massless Dirac-Weyl equation, with Fermi velocity $v_F\approx 
c/300$ instead of the speed of light $c$. If the graphene layer is subject to external magnetic fields 
orthogonal to its surface (the $x-y$ plane), the Dirac-Weyl equation reads:
\begin{eqnarray}
&& \mathbf{H} \Psi(x,y) = \upsilon_{F} \boldsymbol{\sigma}\cdot\left[\mathbf{p}+\frac{e \mathbf{A}}{c}\right] 
\Psi(x,y) = E \Psi(x,y), \label{e2}
\end{eqnarray}
where $v_F \sim 8\times 10^5 m/s$ is the Fermi velocity, $\boldsymbol{\sigma} = (\sigma_x,\sigma_y)$ are
the Pauli matrices, $\mathbf{p} = -i \hbar(\partial_x,\partial_y)^T$ is the momentum operator in the 
$x-y$ plane, $-e$ is the electron charge, and $\mathbf{A}$ is the vector potential leading to the magnetic 
field through $\mathbf{B}= \boldsymbol{\nabla} \times \mathbf{A}$. For magnetic fields which change just 
along $x$-direction, ${\bf B}= \mathcal{B}(x) {\hat e}_z$, in the Landau gauge we have that $\mathbf{A} =  
\mathcal{A}(x) \hat{e}_y, ~\mathcal{B}(x) = \mathcal{A}'(x)$. Since there is a translational invariance 
along $y$ axis, we can propose 
\begin{eqnarray}
\Psi(x,y) = e^{i k y} \left[
                        \begin{array}{c}
                          \psi^+(x) \\
                          i \psi^-(x) \\
                        \end{array}
                      \right],\label{e11}
\end{eqnarray}
where $k$ is the wave number in the $y$ direction and $\psi^\pm(x)$ describe the electron amplitudes on 
two adjacent sites in the unit cell of graphene. Thus we arrive to:
\begin{eqnarray}
&& \left(\pm \frac{d}{dx} + \frac{e}{c \hbar} \mathcal{A} + k \right) \psi^\mp(x) = 
\frac{E}{\hbar \upsilon_F} \psi^\pm(x).
\end{eqnarray}
By decoupling these set of equations it is obtained:
\begin{eqnarray}
&& H^\pm \psi^\pm(x) = \mathcal{E} \psi^\pm(x), \qquad \mathcal{E} = \frac{E^2}{\hbar^2 \upsilon_F^2},  \\
&& H^\pm = -\frac{d^2}{dx^2} + V^\pm =  -\frac{d^2}{dx^2} + \left(\frac{e \mathcal{A}}{c \hbar} + k\right)^2 \pm \frac{e}{c \hbar} \frac{d \mathcal{A}}{dx}.  
\end{eqnarray}
Let us note that these expressions are characteristic of the first-order SUSY QM. In fact, through the
identification\footnote{We choose here a notation consistent with section 2. Please do not confuse the
intertwining operators of equation~($\ref{intergraph}$) with the magnetic field ${\bf B}$, its magnitude
$\mathcal{B}(x)$ or any of its components.}:
\begin{eqnarray}
&& B^\pm = \mp \frac{d}{dx} + {\cal W}(x),  \label{intergraph}
\end{eqnarray}
where 
\begin{eqnarray}
{\cal W}(x) = \frac{e \mathcal{A}(x)}{c \hbar} + k,
\end{eqnarray}
is the superpotential, it turns out that
\begin{eqnarray}
&& B^\mp \psi^\mp(x) = \sqrt{\mathcal{E}} \psi^\pm(x).
\end{eqnarray}
The SUSY partner Hamiltonians $H^\pm$ thus satisfy:
\begin{eqnarray}
& H^\pm = B^\mp B^\pm, \quad
V^\pm(x) = {\cal W}^2 \pm {\cal W}', \\
& H^\pm B^\mp = B^\mp H^\mp.
\end{eqnarray}
By comparing these expressions with the formalism of section~2, one realizes that $H^\pm$ can be 
identified with any of the two SUSY partner Hamiltonians $H$ and $\widetilde H$ (up to a constant 
factor), depending of which one will be taken as the departure Hamiltonian. Moreover, by deriving the 
superpotential with respect to $x$ it is obtained:
\begin{eqnarray}
&& \mathcal{B}(x) = \frac{c\hbar}{e} \frac{d{\cal W}}{dx}.
\end{eqnarray}
This formula suggests a method to proceed further: the magnetic field $\mathcal{B}(x)$ has to be 
chosen cleverly, in order to arrive to a pair of exactly solvable potentials $V^\pm$. In particular, 
it has been chosen in several different ways but taking care that $V^\pm$ are shape invariant potentials 
\cite{knn09}. An important case of this type appears for constant homogeneous magnetic fields: in 
such a situation both $V^\pm$ become harmonic oscillator potentials. It is worth to mention also that 
the shape invariance condition has been generalized, thus supplying a method for generating magnetic 
fields which are deformed with respect to the chosen initial one, but leading once again to an exactly 
solvable problem \cite{mf14}.

Let us note that the SUSY methods have been applied also to other carbon allotropes, as the carbon 
nanotubes, and it has been successfully implemented when electrostatic fields are applied, with or 
without static magnetic fields. In addition, the coherent state methods started to be applied 
recently to graphene subject to static homogeneous magnetic fields \cite{df17}. As can be seen, 
the SUSY methods applied to Dirac materials is a very active field which surely will continue its 
development in the near future \cite{knn09,mtm11,jp12,pr12,jknt13,qs13,mf14,knt15,sr17,chrv18,jss18}.

At this point, it is worth to mention also the applications of SUSY QM to optical system, since there is 
a well known correspondence between Schr\"odinger equation and Maxwell equations in the paraxial 
approximation. Thus, it seems natural to think that many techniques successfully used to deal with 
quantum mechanical problems can be directly applied to optical systems in an appropriate approximation. 
In a way, we are dealing with the optical analogues of quantum phenomena, which have been realized for 
example in waveguide arrays, optimization of quantum cascade lasers, among others. In particular, the 
optical analogues of SUSY QM is an emergent field which could supply a lot of interesting physical 
information \cite{cw94,mhc13,zrfm14,mi14,dllrc15,cg17,mwlf18,cj18}.

\section{Conclusions}

It has been shown that supersymmetric quantum mechanics is a simple powerful tool for generating 
potentials with known spectra departing from a given initial solvable one. Since the spectrum of the new 
Hamiltonian differs slightly from the initial one, the method can be used to implement the spectral 
design in quantum mechanics. 

In this direction, let us note that here we have discussed real SUSY transformations, by employing 
just real seed solutions which will produce at the end real SUSY partner potentials $\widetilde V(x)$. 
However, most of these formulas can be used without any change for implementing complex SUSY 
transformations. If we would introduce this procedure gradually, in the first place we could use complex 
seed solutions associated to real factorization energies in order to generate complex potentials with 
real spectrum \cite{aicd99,bf11b}. This offers immediately new possibilities of spectral design which 
were not available for real SUSY transformations, for example, through a complex first-order SUSY 
transformation with real factorization energies a new energy level can be created at any position on the 
real energy axis. In a second step of this approach, one can use complex seed solutions associated to 
complex factorization energies for an initial potential which is real \cite{fmr03}, thus generating new 
levels at arbitrary positions in the complex energy plane. The third step for making complex the SUSY 
transformation is to apply the method to initial potentials which are complex from the very beginning 
\cite{fg15}. In all these steps we will get at the end new potentials which are complex, but the 
spectrum will depend of  the initial potential as well as of the kind of seed solutions employed. 

We want to finish this paper by noting that the factorization method and intertwining techniques have 
been also applied with success to some discrete versions of the stationary Schr\"odinger equation 
\cite{go05,do06,dod07,dj07,dj17,df18}. The connections that could be established between such  problems 
and well known finite difference equations \cite{aac07,nsu91} could contribute to the effort of 
classifying the known solutions and generate new ones, as it has happened in the continuous case for 
more than eighty years.

As it was pointed out previously, one of our aims when writing this article was to make a short review 
of the most recent advances of SUSY QM, either on purely theoretical or applied directions. We hope to 
have succeeded; perhaps the reader will find interesting and/or useful the ideas here presented. 

\bigskip

\noindent{\large\bf Acknowledgments.}
The author acknowledges the financial support of the Spanish MINECO (project MTM2014-57129-C2-1-P) and
Junta de Castilla y Le\'on (VA057U16).

\end{document}